\documentclass[%
 reprint,
superscriptaddress,
showpacs,preprintnumbers,
twocolumn,
 amsmath,amssymb,
 aps,
prb,
]{revtex4-1}
\usepackage{graphicx}
\usepackage{csquotes}
\usepackage{bm}
\usepackage{color}
\usepackage{soul}

\begin{document}

\title{Accurate formation energies of charged defects in solids: a systematic approach}

\author{Dmitry Vinichenko}
\affiliation{Department of Chemistry and Chemical Biology, Harvard University, Cambridge, Massachusetts 02138, USA}
\author{Mehmet Gokhan Sensoy}
\affiliation{John A. Paulson School of Engineering and Applied Sciences, Harvard University, Cambridge, Massachusetts 02138, USA}
\affiliation{Department of Physics, Middle East Technical University, Ankara, 06800, Turkey}

\author{Cynthia M. Friend}
\affiliation{Department of Chemistry and Chemical Biology, Harvard University, Cambridge, Massachusetts 02138, USA}
\affiliation{John A. Paulson School of Engineering and Applied Sciences, Harvard University, Cambridge, Massachusetts 02138, USA}
\author{Efthimios Kaxiras}
\email{kaxiras@physics.harvard.edu}
\affiliation{John A. Paulson School of Engineering and Applied Sciences, Harvard University, Cambridge, Massachusetts 02138, USA}
\affiliation{Department of Physics, Harvard University, Cambridge, Massachusetts 02138, USA}

\date{\today}

\begin{abstract}
Defects on surfaces of semiconductors have a strong effect on their reactivity and catalytic properties. The concentration of different charge states of defects is determined by their formation energies. First-principles calculations are an important tool for computing defect formation energies and for studying the microscopic environment of the defect. The main problem associated with the widely used supercell method in these calculations is the error in the electrostatic energy, which is especially pronounced in calculations that involve surface slabs and 2D materials. We present an internally consistent approach for calculating defect formation energies in inhomogeneous and anisotropic dielectric environments, and demonstrate its applicability to the cases of the positively charged Cl vacancy on the NaCl (100) surface and the negatively charged S vacancy in monolayer MoS$_{\rm{2}}$.
\end{abstract}

\maketitle

\section{Introduction}
Defects play an important role in the electronic and structural properties of semiconductors, so understanding of defect behavior is critical for materials design.~\cite{RN486, RN108, RN105, RN100} The most important quantity for a given defect type is the formation energy, since it determines the concentration of the defect in the material. Density functional theory (DFT) based calculations provide unmatched insight into defect formation energies and defect microscopic structure~\cite{RN525,RN98} which can complement a number of experimental techniques for studying defect properties, ranging from scanning tunneling microscopy to electron paramagnetic resonance.~\cite{RN529,RN531,RN530,RN533,RN535,RN538,RN540,RN532} In DFT calculations, the widely used supercell method is capable of addressing structural changes in the material but suffers from systematic errors when dealing with charged defects, due to the use of periodic boundary conditions. This constraint makes necessary the introduction of an implicit neutralizing background charge, which adds spurious terms to the total energy of the system.~\cite{RN487, RN93, RN556,RN97,RN28,RN559,RN92,RN555} A number of methods for addressing this problem have been proposed, but most of them are not applicable to supercells with variable and anisotropic dielectric profile. The simplest of corrections accounting for electrostatic interaction is the Makov-Payne correction, \cite{RN1000} amounting to a difference between electrostatic energy of a point charge under open boundary conditions, and the Madelung sum for its energy under periodic boundary conditions.  However, in practical applications it has been proven hard to use this correction reliably,\cite{RN1001,RN1002} the main reason being that the expression for the correction energy has the macroscopic dielectric constant in the denominator but the supercell method deals with the material on a microscopic scale and therefore the bulk limit might not be applicable. Accordingly, alternative schemes were developed to calculate the true formation energy of an isolated defect for a series of supercells with the same shape and progressively increasing size,\cite{RN92,RN555,RN24} followed by fitting to a scaling law with the inverse size of the supercell while treating the dielectric constant of the material as a parameter of the model; variants of the scheme accounting for anisotropic dielectric tensor have been also implemented.\cite{RN1003} Recent works aimed at addressing this issue have concentrated on treating strictly two-dimensional materials.~\cite{PRX40310442014,PRL1141968012015} A method for correcting the charged defect formation energies was introduced by Freysholdt, Neugebauer and Van de Walle (FNV) through alignment of the defect-induced potential using the planar-averaged electrostatic potential without including relaxation.\cite{RN556} However, the defect-induced potential is significantly affected by atomic relaxation which reduces the accuracy of the calculations based on this approach. Moreover, this scheme requires substantial computational effort for calculating the correction energy, since it relies on calculating the supercells both with and without the presence of defects. Recently, the method by Kumagai \emph{et al.} \cite{RN559} proposed to correct the defect formation energy by extending the FNV scheme by using the atomic site potential. In this study, it was also shown that the potential alignment can be eliminated for the defect formation energies. Using the atomic site potential in this method is not efficient in small supercells and gives rise to non-negligible sampling errors. Komsa \emph{et al.}\cite{RN27} proposed a method for correcting the electrostatic energy of charged defects which obtains the charged defect formation energy in 3D materials by estimating the electrostatic energy of localized charged defects and the neutralizing charge in a dielectric environment. This method is not applicable for 2D systems and needs to re-construct the dielectric constant profile of the system.\cite{PRX40310442014} Previously proposed methods mostly rely on a combination of two procedures: (i) modeling the electrostatic energy of the defect-induced charge which is the standard definition in the literature~\cite{RN556} and (ii) employing the concept of potential alignment. In most methods the computation of the electrostatic energy is typically implemented for defects in the bulk and with a simplifying model for the defect-induced charge.~\cite{RN93} The potential alignment term is due to the use of a model, typically a Gaussian, for the defect charge~\cite{RN559} and, as we demonstrate, can be eliminated altogether. In this work, we present a systematic and consistent approach to computing charged defect formation energies in complex dielectric environments and we provide guidance on its computationally efficient implementation. In addition, we highlight some important technical details of the calculation procedure such as the appropriate mode of extrapolation of the energy computed under periodic boundary conditions and the trimming process to make the model supercell for electrostatic calculations. Most importantly, we show that the proposed method allows us to treat both cases, bulk 3D materials and 2D materials embedded in vacuum, on the same footing, as well as to include relaxation of the ions.

\section{Method Description}
The method we propose for calculating the true formation energy of a charged defect, $E_f(q)$, is a post-processing correction to the total energy of the supercell with a charged defect obtained from DFT, $E_{\rm{DFT}}^{def}$($q$):

\setlength{\parindent}{0pt}
\begin{multline}
E_f(q)=E_{\rm{DFT}}^{def}(q)-E_{\rm{DFT}}^{st}+\sum_{i}\mu_{i}n_{i}\\
+q(E_{\rm{VBM}}+E_{F})+E_{corr}
\end{multline}\label{eqn:eq1}

where E$_{\rm{DFT}}^{st}$ is the DFT total energy of the stoichiometric slab, $\mu_i$ the chemical potentials of the species added or removed to create the defects under appropriate thermodynamic conditions, $n_i$ the stoichiometric coefficients for those species, $E_{\rm{VBM}}$ the valence band maximum energy, $E_F$ the Fermi level with respect to valence band maximum, and $E_{corr}$ the correction energy in our method. The supercell model for charged defects implicitly imposes a compensating background charge to make the supercell overall neutral. This model of an infinite array of defects immersed in the background charge is very different from the target, that is, an isolated defect in the host material. As was shown before,~\cite{RN486,RN27} the difference in total energy between those models can be captured by an energy correction, $E_{corr}$, which involves subtracting the electrostatic energy of the incorrect model, $E_{\rm{PBC}}$ and adding the electrostatic energy of the isolated defect-induced charge, $E_{iso}$. In the following discussion, we explain the correction method for the case of a charged chlorine vacancy, V$_{\mathrm{Cl}}^+$ on the NaCl (100) surface to facilitate comparison to previously proposed methods. We also discuss the applicability of the method to 2-dimensional materials (such as graphene, BN, or MoS$_{\rm{2}}$) by considering the case of the charged sulfur vacancy, V$_{\mathrm{S}}^-$ in MoS$_{\rm{2}}$.
For DFT computations we use the QuantumEspresso package.~\cite{RN423} For the simulation of NaCl surfaces we use a 2~$\times$~2~$\times$~3 supercell with 4~$\times$~4~$\times$~1 k-point sampling grid, kinetic energy cutoffs for plane-wave expansion of the wavefunctions equal to 30 Ry and of the density equal to 300 Ry. For MoS$_{\rm{2}}$ we use a 6~$\times$~6 supercell that can be cast into a rectangular shape, a vacuum region of size 16~\AA, with $\Gamma$ point sampling of the Brillouin zone, and kinetic energy cutoffs equal to 50 Ry for the wavefunctions and 500 Ry for the charge density.
\setlength{\parindent}{12pt}
\section{Electrostatics under periodic boundary conditions}

The computation of $E_{\rm{PBC}}$ is based on solving the Poisson equation under periodic boundary conditions for the electrostatic potential $V_{\rm{PBC}}(\vec{r})$:

\setlength{\parindent}{0pt}
\begin{equation}
\epsilon_0\nabla [\epsilon(z)\nabla V_{\rm{PBC}}(\vec{r})]=-\rho_d (\vec{r})
\end{equation}\label{eqn:eq2}

where $\epsilon_0$ is the vacuum permittivity, $\epsilon(z)$ is the dielectric profile of the model slab in the direction perpendicular to the surface (this can be extended to anisotropic materials, as discussed in Appendix A), and $\rho_d(\vec{r})$=$\vert \varphi(\vec{r}) \vert^2$ is the charge induced by the defect level in the band gap. The incorrect electrostatic energy can be computed by integration over the supercell volume:

\begin{equation}
E_{\rm{PBC}}=\frac{1}{2}\int \rho_d(\vec{r})V_{\rm{PBC}}(\vec{r})d\vec{r}
\end{equation}\label{eqn:eq3}

\setlength{\parindent}{12pt}
This model has two key parameters: the defect charge $\rho_d(\vec{r})$ and the shape of the dielectric profile $\epsilon(z)$. The main contribution of our approach is a consistent treatment of the electrostatic model computation.

Instead of using a Gaussian distribution for the defect-related charge, we use the actual $\vert \varphi(\vec{r}) \vert^2$ obtained from the DFT calculation. A Gaussian model is often used due to the availability of analytical expressions for the electrostatic energy and fast convergence of the model electrostatic energy with respect to the discretized mesh size. In our parallel implementation of the potential computation, this is not an important factor and we can explicitly use the defect wavefunction in the Poisson equation. There are several reasons for doing this: first, we find that often the corresponding defect wavefunctions are highly anisotropic and have several lobes (Fig.~1), so a smooth Gaussian model is an inappropriate description; second, the complex shape of the wavefunction leads to a substantial ambiguity in locating the center of the Gaussian and our model calculations reveal that a shift of the 
\begin{figure}
\centering
\includegraphics[scale=0.55]{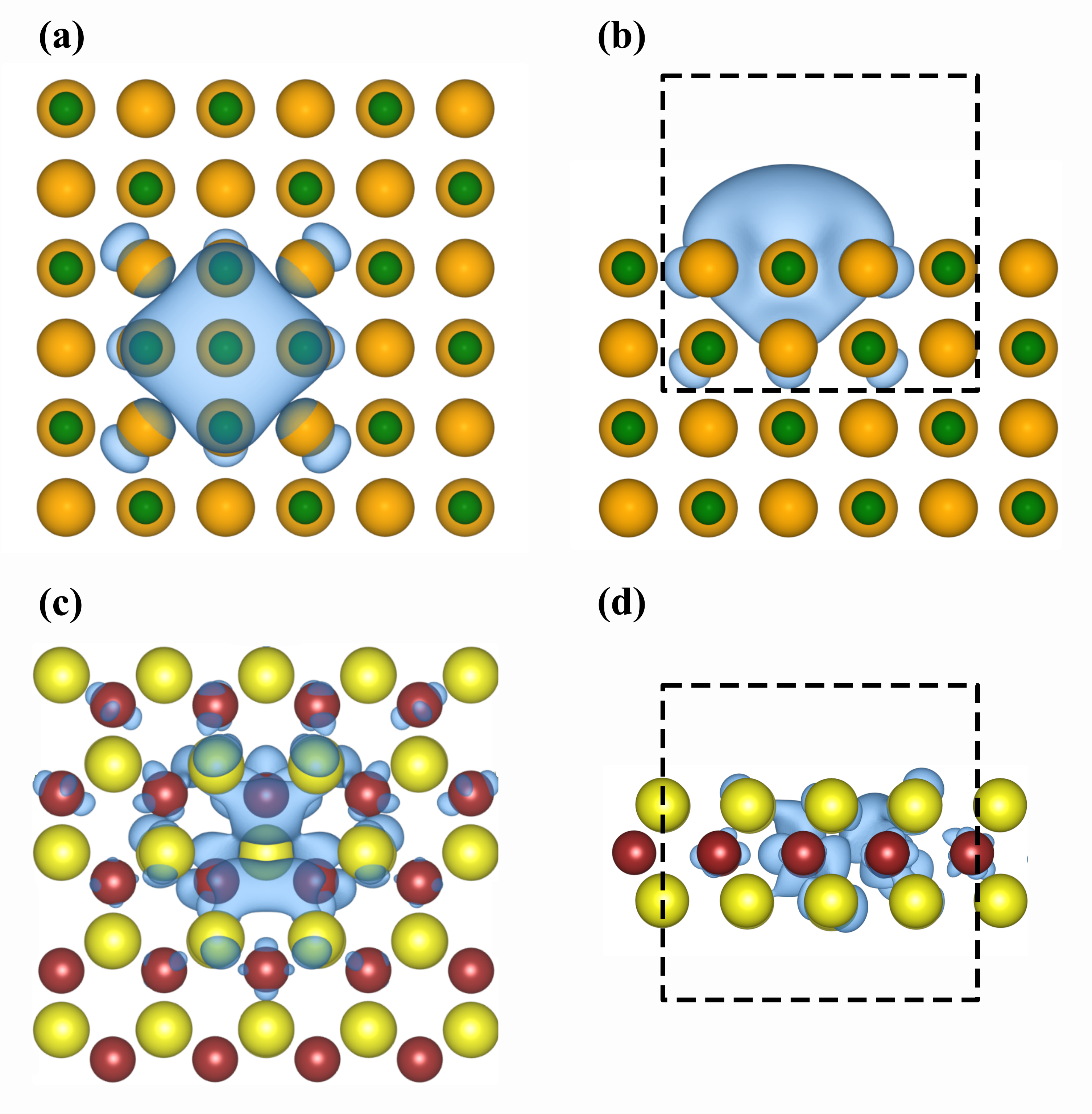}
\caption{Defect charge distributions. (a) Top view of the NaCl (100) surface with V$_{\mathrm{Cl}}^+$: green --Na ions, orange--Cl ions. The blue cloud represents the $\rho_d(\vec{r})$ = $\vert \varphi(\vec{r}) \vert^2$ for the defect level in the bandgap. (b) Side view of the same surface slab, showing the slight asymmetry in the charge distribution shape and additional lobes on Cl atoms around the vacancy. The dashed line denotes the trimmed cubic part of the defect charge used in the extrapolation procedure. (c) Top view of MoS$_{\rm{2}}$ monolayer with V$_{\mathrm{S}}^-$ : yellow--S ions, red--Mo ions. The complex multi-lobe structure of the charge distribution is apparent. (d) Side view of the V$_{\mathrm{S}}^-$ charge distribution: the difference in spatial extent in-plane and out-of-plane is apparent.}
\end{figure}\label{fig:fig1}charge center in the direction perpendicular to the surface by 0.15 Bohr (well within the ambiguity involved) results in changing the electrostatic model energy by 0.1 eV. Similarly, the anisotropic shape of the wavefunction results in a poorly determined Gaussian width and the uncertainty in this parameter leads to differences of up to 0.25 eV in the model energy. Moreover, for multi-lobe defect wavefunctions, like those related to forming a sulfur vacancy in MoS$_{\rm{2}}$, the over-estimate in the width of the Gaussian can lead to  \enquote{spilling over} of the model charge from the simulation cell, which is the case when the cell dimensions are smaller than 8 standard deviations of the Gaussian ($\pm$4$\sigma$ is required to contain 99.99\% of the charge). This is important, since we find that losing more than 0.1\% of the charge results in errors in electrostatic energy on the order of 0.1 eV. 

The other input to our method is the shape of the dielectric profile. For this, we use a model of two constant dielectric regions joined by error functions at the interfaces; the parameters defining the profile are: the material's dielectric constant in each region (for vacuum it is 1.0 by definition) and the positions of the interfaces. Previous work suggests obtaining the dielectric profile from the DFT calculations, for example from the response of the model slab to an applied electric field.~\cite{RN27} This is not necessary for the following reasons: first, the DFT simulations have intrinsic limitations due to the commonly employed semi-local exchange-correlation functionals, and fail to reproduce the experimental values of the dielectric constant; second, in this model we seek to capture the response of the semiconductor to the defect charge at the microscopic level, and the value of the bulk experimental dielectric constant is not necessarily optimal for it; third, this method ceases to be applicable when the ionic relaxations are included, because then the field-induced ionic displacements result in substantial rearrangements of the electronic density, leading to discontinuities in the dielectric profile. Instead, we model the dielectric profile approximately using the experimental value for the dielectric constant of the material, as a starting point, and the average of atomic radii of the surface atoms to get the profile boundaries; as shown below for the case of 2D materials, the electrostatic correction is fairly insensitive to the value of the dielectric constant, which makes the use of the experimental value as a starting point perfectly reasonable. We then fine-tune those parameters in order to achieve alignment to the DFT potential, as described next.

The notion of potential alignment defines the mismatch of the potentials induced by the unscreened defect charge in the model calculation and in the actual DFT computation.~\cite{RN559} This term is typically expressed as 

\setlength{\parindent}{0pt}
\begin{equation}
\Delta V=V_{\rm{PBC}} \vert_{far}-[V_{\rm{DFT}}^{st}-V_{\rm{DFT}}^{def}(q)]\vert_{far}
\end{equation}\label{eqn:eq4}

where $V_{\rm{DFT}}^{st}$ is the electrostatic potential for the stoichiometric slab, $V_{\rm{DFT}}^{def}(q)$ is the potential for the slab with a charged defect and the subscript \textit{\enquote{far}} denotes the vacuum region of the supercell farthest from the defect. The potential alignment term arises from the approximations made in the electrostatic model. The main difference between the used method here from earlier methods \cite{RN28,RN27} in that our approach eliminates the potential alignment term, by modeling the electrostatic environment of the simulation cell. Since we are using the exact wavefunction of the defect, we adjust the dielectric profile parameters in a way that properly aligns the model potential and the DFT potential difference in the vacuum region of the simulation supercell far from the defect.
\setlength{\parindent}{12pt}

The model electrostatic potential has qualitatively different dependence on the dielectric constant and the positions of the profile boundaries, as shown in Fig.~2: varying the value of the dielectric constant changes the amplitude of the features on the model potential and the slope in the alignment region. For NaCl we choose the value of 2.8 which results in flat $\Delta$V, see Fig.~2. Variation of the profile boundary position results in a rigid shift of the potential in vacuum. Overall, by adjusting those parameters one can find a combination resulting in a flat line close to zero for $\Delta V$ denoted by the black circles on Fig.~2 in the region far from the defect position. It is important to emphasize that the presence of the $\Delta$V term is solely due to the use of a crude model for the defect-induced charge $\vert \phi d(\vec{r}) \vert ^2$; it has been shown that the potential alignment term is unnecessary if the electrostatic energy part of the problem is properly described.\cite{RN93,RN556,RN97} Another motivation to remove the potential alignment term is the fact that it becomes increasingly hard to define it when the relaxation of ionic positions in the material are included in the model, since the displacement of atoms changes the electrostatic potential substantially, and it becomes practically impossible to carry out the alignment with the far-field bulk-like region in the expression for $\Delta$V.

\begin{figure}
\centering
\includegraphics[scale=0.5]{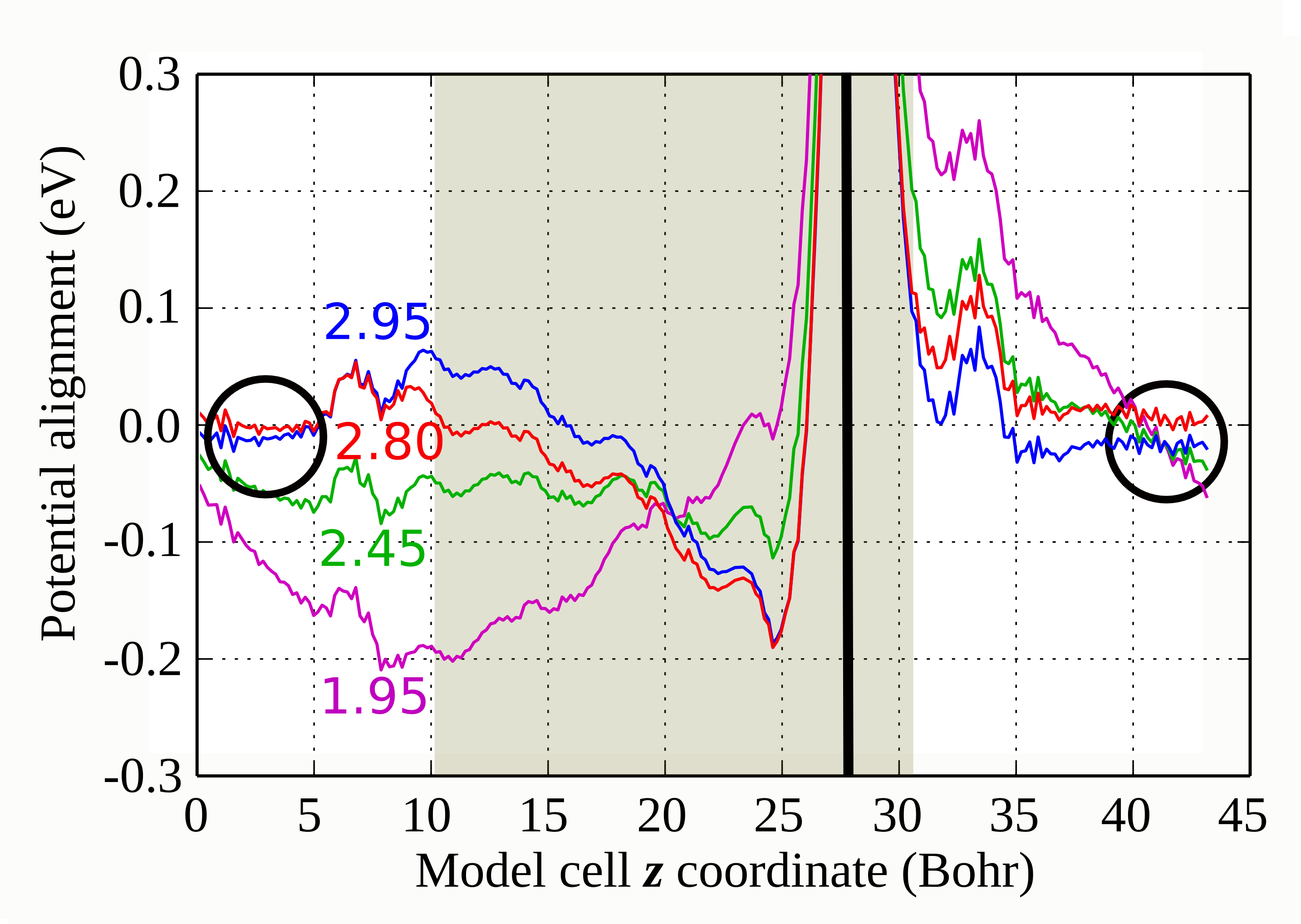}
\caption{Potential alignment procedure. The colored lines represent \textit{xy} plane-averaged values of the potential alignment $\Delta V$ defined in Eq.(4) for the NaCl slab, with the V$_{\mathrm{Cl}}^+$ located at \textit{z} = 28 Bohr. The circle denotes the \textit{\enquote{far}} region used for alignment. The legend denotes the value of the dielectric constant of NaCl used in the construction of the model dielectric profile. The shaded area denotes the region occupied by the material.}
\end{figure}\label{fig:fig2}

Finally, we note that the inaccuracies associated with sampling the defect-induced charge lead to errors of about 0.03 eV in the values of $E_{\rm{PBC}}$. These errors do not converge fast with finer mesh sampling, so there is no need to specifically increase the sampling and plane wave expansion cutoff in the DFT calculations. Moreover, we find that downsampling the output wavefunction by a factor of 2 or 3 (so that the mesh size is about 0.3--0.4 Bohr) changes the $E_{\rm{PBC}}$ by about 1 meV, which can be used to choose computational parameters optimally to reduce the cost of the calculations.

For an isolated charge the boundary conditions in the Poisson equation are $\displaystyle{\lim_{\vec{r} \to \infty} V(\vec{r}) = 0}$, which requires infinitely large simulation domain. A proper way to treat this condition is to perform a direct pairwise summation of interaction energies for discretized charge elements, including two different dielectric media through the image charge method (Appendix B). This approach is computationally intensive (it scales as the sixth power of the mesh size) and allows only one sharp boundary, a rather severe approximation to the real material interface.

A different approach is extrapolation of the energy under periodic boundary conditions to the limit of the infinite cell size. The dependence of the model $E_{\rm{PBC}}$ on the inverse cell size is linear, which allows easy extrapolation,~\cite{RN27} with the only parameter being the factor by which the model cell is extended in all dimensions. We analyzed this method for a model system of Gaussian charge in vacuum, as well as for real materials; we find that in practice a maximal scale of 5 can be used, resulting in extrapolation errors below 0.03 eV for charges in vacuum, and smaller errors for real materials. The electrostatic correction decreases with the increase in the size of the original supercell, so for larger systems the error is dominated by DFT errors in the $E_{\rm{PBC}}$ values, which depend on the sampling of the defect charge state, especially for anisotropic wavefunctions, and can reach 0.03 eV.

\begin{figure}
\centering
\includegraphics[scale=0.8]{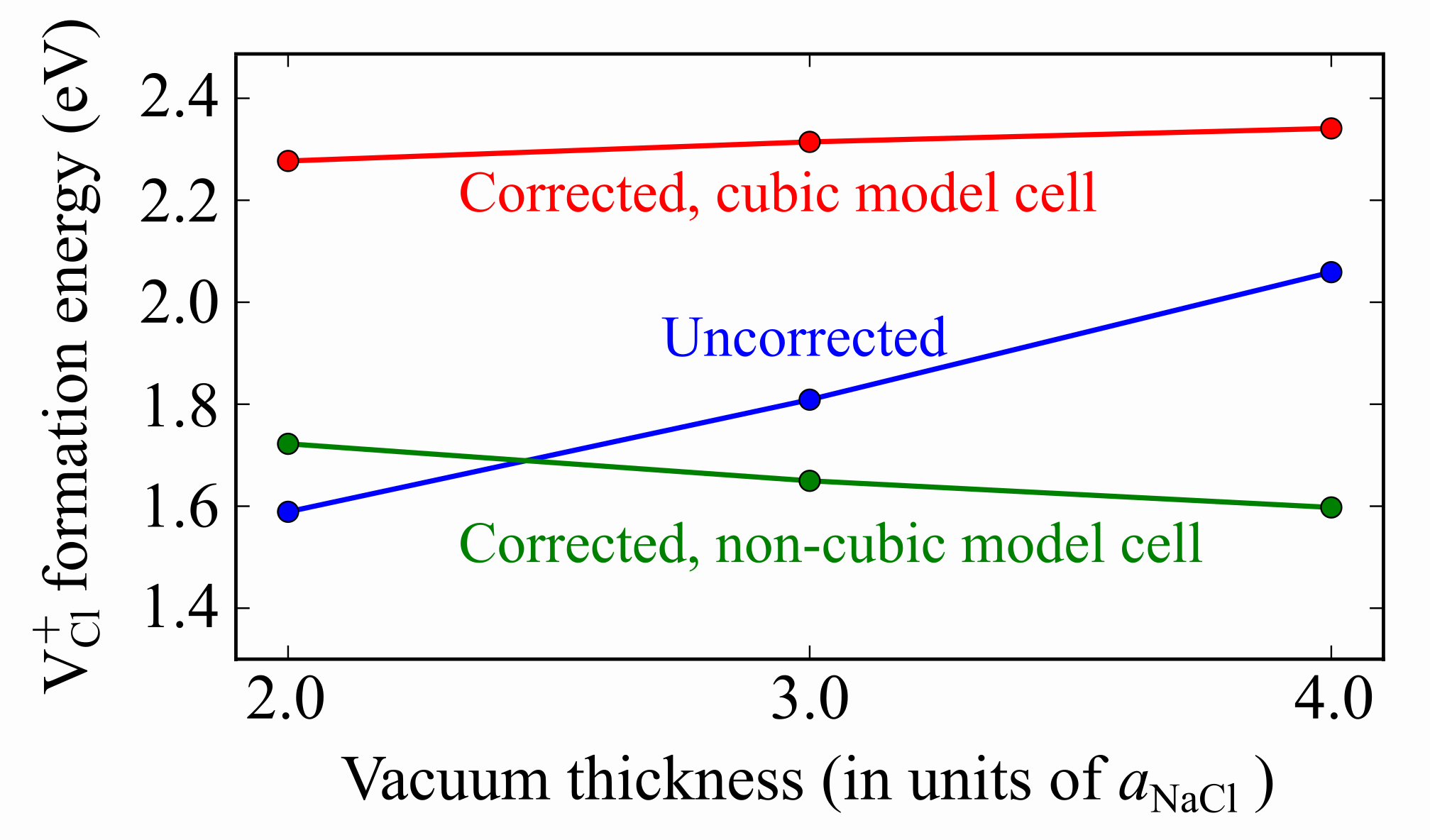}
\caption{Formation energy of V$_{\mathrm{Cl}}^+$ on the NaCl (100) surface as a function of vacuum size (in units of a$_{\mathrm{NaCl}}$ = 10.6 Bohr). Uncorrected (blue) and corrected energies with non-cubic model cell (green) used for extrapolation show large variance; the correct extrapolation procedure gives consistent formation energy values within 0.06 eV (red).}
\end{figure}\label{fig:fig3}

Another important aspect of the problem is that the extrapolation is valid only for a model supercell of strictly cubic shape; extrapolation from cells of different shape result in vastly different and incorrect $E_{iso}$ values. Accordingly, when simulating real materials, the defect wavefunction has to be trimmed to a cubic shape (see Figs. 1(b), 1(d)) for use in the extrapolation procedure. Specifically, upon scaling the system, we pad the trimmed charge distribution with zeros on all sides, placing it in the center of the scaled cell. For the case of semiconductor surface regions, we calculate the position of the dielectric profile boundary closest to the charge by setting the offset to be the same as in the original cell. The second boundary position is calculated by scaling the thickness of the material proportionally to the supercell size.

We investigate the performance of the correction scheme by calculating formation energies of V$_{\mathrm{Cl}}^+$ on the NaCl (100) surface for several supercells with varying vacuum thickness and lateral dimensions. The results are shown in Fig.~3 for the case of varying vacuum thickness. 
\begin{figure}
\centering
\includegraphics[scale=0.95]{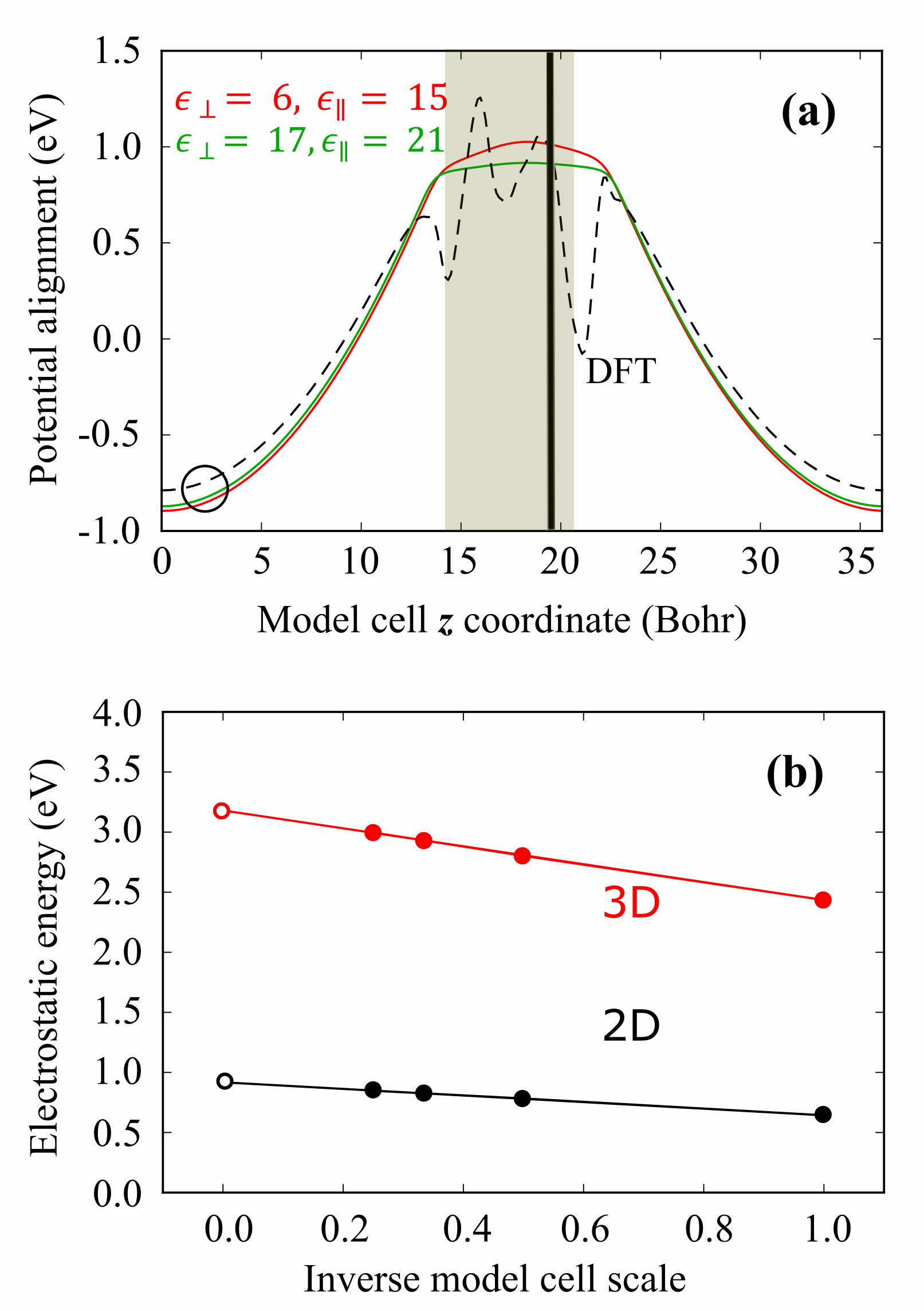}
\caption{Electrostatics for MoS$_{\rm{2}}$: (a) The difference of DFT potentials for V$_{\mathrm{S}}^-$ (black, dashed) and model potentials for a variety of choices for in-plane and out-of-plane components of the dielectric tensor. The evident mismatch in the alignment region can be fixed by moving the positions of the dielectric profile boundaries outwards. (b) The extrapolation procedure illustrated for the cases of scaling the material thickness in the model profile, as for NaCl (\enquote{3D}) and of keeping constant the material thickness, as for MoS$_{\rm{2}}$ (\enquote{2D}). Both cases show a linear dependence on the inverse scale of the model cell.}
\end{figure}\label{fig:fig4}The variance in the uncorrected energies (blue line) is as substantial as the variance in corrected energies with extrapolation from the wavefunction charge distributions of non-cubic shape (green line). Only correction with the proper extrapolation procedure gives consistent formation energies within 0.06 eV, independent of the supercell shape. Analogously, the dependence on the lateral size of the cell is eliminated.

It is important to note that there are two different simulation cells: the one used in DFT, and the one used for the electrostatic computation. The latter one is obtained by discretizing the defect-related wavefunction and casting it to a cubic shape (\enquote{trimming}). The trimming procedure is introduced to make the model supercell for electrostatic calculations cubic, since only in that case the extrapolated isolated boundary conditions energy is correct (see Fig. 3). Since the Poisson equation is solved in Fourier space, the exact position of the charge inside the simulation cell is immaterial, as long as it is approximately in the center of the cell, and in order to achieve that, the model charge is translated to the middle of the model cell for the electrostatic computation. The size of the trimmed supercell is chosen as the smallest one possible in the DFT supercell.

\section{Application to 2-dimensional materials}

The above scheme can be successfully used for 2D materials as demonstrate with the example of the V$_{\mathrm{S}}^-$ defect in a MoS$_{\rm{2}}$ monolayer. The only change needed is the method of scaling the model profile in computing $E_{iso}$: in this case, the positions of both profile boundaries are fixed relative to the charge, which results in keeping the material thickness constant throughout the extrapolation procedure. An important feature of low-dimensional systems is that the actual values of the diagonal elements of the dielectric tensor do not affect the model potential as much as the positions of the boundaries of the dielectric profile. As shown in Fig.~4, the values of the model potential in the alignment region are very close. We use the values of $\epsilon_\perp$ = 6 for the out-of-plane component and $\epsilon_\parallel$ = 15 for the in-plane component; we find the optimal position of the profile boundaries to be at an offset of 2.7 Bohr outwards from the S atoms. The dependence of $E_{\rm{PBC}}$ on the inverse scale of the model cell is similarly linear, as shown in Fig.~4. Application of our correction scheme results in elimination of the spurious dependence of the vacancy formation energy on the vacuum layer thickness, the corrected formation energies being consistent to within 0.06 eV.

\section{Summary}
To summarize, we presented an internally consistent scheme for computation of charged defect formation energies in systems with complex dielectric profiles. The overall algorithm is the following:
\begin{enumerate}	
\item{Construct the stoichiometric and defected slabs, obtain $\vert \varphi(\vec{r}) \vert^2$, the defect charge density, the level of the VBM, and electrostatic potentials $V_{\rm{DFT}}^{st}$, $V_{\rm{DFT}}^{def}(q)$.}
\item{Fine-tune the parameters of the model dielectric profile, that is, the values of the dielectric constant and the positions of interfaces in order to achieve alignment between the model $V_{\rm{PBC}}$ and $V_{\rm{DFT}}^{st}$ -- $V_{\rm{DFT}}^{def}(q)$; calculate the corresponding $E_{\rm{PBC}}$.}
\item{Trim $\vert \varphi(\vec{r}) \vert^2$ to a cubic shape, change the dielectric boundary positions accordingly, calculate $E_{\rm{PBC}}$ for a series of scaled model cells; obtain $E_{iso}$ through extrapolation to infinite cell size.}
\item{Add the correction $E_{corr}$ = $E_{iso}$ -- $E_{\rm{PBC}}$ to the defect formation energy.}
\end{enumerate}
We find that the electrostatic correction described here is best suited for applications to 2D materials or semiconductors with low ($<$ 10) dielectric constant. In materials with stronger screening the value of the electrostatic correction is small; at the same time, introduction of charged defects into the supercell results in substantial rearrangements of atoms, which are hard to contain in a supercell, even of a size as large as 1000 atoms. This leads to large errors due to elastic energy contributions, which become the dominant term among errors associated with the supercell method for such materials (an example of such a case is TiO$_{\rm{2}}$).

\renewcommand{\thetable}{A\arabic{table}}
\renewcommand{\thefigure}{A\arabic{figure}}
\renewcommand{\thesection}{A\arabic{section}}   

\appendix

\section{Energy with periodic boundary conditions}

The computational scheme described here is based partly on a previous work proposing a method for defect formation energy computations.~\cite{RN27} In Fourier space the Poisson equation, Eq.~(2), takes the form

\setlength{\parindent}{0pt}
\begin{equation}
\widehat{\epsilon} (G_z)\ast\vert G \vert^2 \widehat{V}(\vec{G}) + G_z \widehat{\epsilon}(G_z) \ast G_z\widehat{V}(\vec{G}) = \widehat{\rho _d}(\vec{G}) 
\end{equation}\label{eqn:eq5}

where $\widehat{\epsilon}$, $\widehat{V}$ and $\widehat{\rho_d}$ are the Fourier transforms of the dielectric profile, the potential and the defect charge, respectively.
In actual computational applications the quantities described above, the charge density of the defect $\rho _d(\vec{r})$ and the corresponding potential $V(\vec{r})$, are represented on a discrete mesh of size ($N_x$, $N_y$, $N_z$), and corresponding mesh spacings $\Delta x$ = $L_x$/$N_x$. With the definition of the mesh in Fourier space, the integral in the convolutions is reduced to a sum, and then the discretized form of the Poisson equation can be simplified as follows:

\begin{multline}
\widehat{\epsilon} (G_z)\ast\vert G \vert^2 \widehat{V}(\vec{G}) + G_z \widehat{\epsilon}(G_z) \ast G_z\widehat{V}(\vec{G}) \\
 = \sum_{G^{\prime}_z} \widehat{\epsilon}(G_z - G^{\prime}_z) G^{\prime}_z{}^2 \widehat{V}(G_x,G_y,G^{\prime}_z)\quad \quad \quad \quad \quad \quad \\
 + \sum_{G^{\prime}_z} \widehat{\epsilon}(G_z- G^{\prime}_z)(G_x^2 + G_y^2)\widehat{V}(G_x,G_y,G^{\prime}_z)\quad \quad \quad\\
 + \sum_{G^{\prime}_z} \widehat{\epsilon}(G_z- G^{\prime}_z)(G_z- G^{\prime}_z)\widehat{V} (G_x,G_y,G^{\prime}_z)G^{\prime}_z\quad \quad \\
\quad = \sum_{G^{\prime}_z} \widehat{\epsilon}(G_z- G^{\prime}_z)(G_x^2 + G_y^2 + G_z G^{\prime}_z)\widehat{V}(G_x,G_y,G^{\prime}_z)\\
 = \widehat{\rho_d}(G_x,G_y,Gz)\quad \quad \quad \quad \quad \quad \quad \quad \quad \quad \quad \quad \quad \\
\end{multline}

which in discrete representation reads:

\begin{equation}
\sum_{l}\epsilon_{k-l+1}[ (G_x^i)^2 + (G_y^j)^2 + G_z^k G_z^l ]V_{ijl} = \rho_{ijk}
\end{equation}\label{eqn:eq8}

where we have introduced the shorthand notation $\epsilon_{k-l+1}$ =$\widehat{\epsilon}$($G_z^k$ -- $G_z^l$ ), $V_{ijl}$ = $\widehat{V}$($G_x^i$,$G_y^j$,$G_z^l$), and $\rho_{ijk}$= $\widehat{\rho_d}$($G_x^i$,$G_y^j$,$G_z^l$). The presence of a non-trivial dielectric profile in the \textit{z} direction results in coupling between components of $V_{ijl}$ and $\rho_{ijk}$ for $k,l$=1...$N_z$. The problem is factorized into $N_x$ $\times$ $N_y$ systems of linear equations defined by matrices $\textbf{M}^{ij}$ with matrix elements $M_{kl}^{ij}$

\begin{equation}
M_{kl}^{ij} = \epsilon_{k-l+1}[ (G_x^i)^2 + (G_y^j)^2 + G_z^k G_z^l ]
\end{equation}\label{eqn:eq9}

\setlength{\parindent}{12pt}
The matrix elements $M_{kl}^{ij}$ can be expressed through the circulant formed from the vector of Fourier components of the dielectric profile, $\widehat{\textbf{C}}$[$\epsilon$]:

\begin{equation}
M_{kl}^{ij} = \widehat{C}_{kl}[\epsilon][ (G_x^i)^2 + (G_y^j)^2] + \widehat{C}_{kl}[\epsilon] G_z^k G_z^l
\end{equation}\label{eqn:eq10}

The second term in the sum is a Hadamard product of the circulant $\widehat{\textbf{C}}$[$\epsilon$] with the matrix \textbf{G} whose matrix elements are defined by $G_{kl}$ = $G_z^k$ $G_z^l$. In modern software libraries the enumeration of wavevectors inside the {G$_z^i$ } set is implemented with the first half of the set being the wavevectors from $G_z^1$ = 0 to $G_z^{N_z /2 +1}$ = $\frac{\pi N_z}{L_z}$ , and the second half of the set (the negative wavevectors in ascending order) from $G_z^{N_z /2 +2}$ = --$\frac{\pi(N_z-1)}{L_z}$ to $G_z^{N_z}$~=~--$\frac{\pi}{L_z}$. With that notation, the outer product matrix \textbf{G} has zero matrix elements along the first row and first column, having rank of $N_z$ -- 1. Therefore, the Hadamard product $\widehat{\textbf{C}}$[$\epsilon$]\textbf{G} also is rank-deficit. For this reason, for the case $i$ = 1, $j$ = 1, when the components $G_x^1$ = 0, $G_y^1$ = 0, so is the first term in the equation above, and the matrix \textbf{M}$^{11}$ is rank-deficient. The component at the head of this matrix establishes the relation between the average value of the charge over the simulation cell, $\rho_{111}$, and the cell average of the electrostatic potential under periodic boundary conditions, $V_{111}$. This can be alleviated by setting $M_{11}^{11}$ equal to an arbitrary number and then setting $V_{111}$ to 0 in the resulting solution.
The scheme described here can be easily extended to the case of the host material with anisotropic dielectric tensor, when instead of one dielectric profile $\epsilon$($z$) the problem will have three profiles corresponding to the components of the dielectric tensor, $\lbrace \epsilon _{xx}$($z$),$\epsilon _{yy}$($z$),$\epsilon _{zz}$($z$)$\rbrace$. After discretization the expressions for matrices \textbf{M}$^{ij}$ can be written in terms of circulant matrices $\widehat{\textbf{C}}$[$\epsilon _{xx}$], $\widehat{\textbf{C}}$[$\epsilon _{yy}$], $\widehat{\textbf{C}}$[$\epsilon _{zz}$] generated from the discrete Fourier transforms of {$\epsilon _{xx}$($z$), $\epsilon _{yy}$($z$), $\epsilon _{zz}$($z$)}, respectively:

\begin{multline}
M_{kl}^{ij} = \widehat{C}_{kl}[\epsilon_{xx}](G_x^i)^2 + \widehat{C}_{kl}[\epsilon_{yy}](G_y^j)^2 \\
+ \widehat{C}_{kl}[\epsilon_{zz}]G_z^k G_z^l\quad \quad \quad \quad \quad \quad \quad \quad \quad \quad
\end{multline}\label{eqn:eq11}

This approach has computational complexity of O($N_x$$N_y$$N_z^{2.8}$) due to $N_x$ $\times$ $N_y$ linear systems of size $N_z$ $\times$ $N_z$. It naturally lends itself to parallelization by distributing the workload for linear systems solution between the processes and then collecting the resulting components of the Fourier transform of the potential.

\renewcommand{\thefigure}{B\arabic{figure}}
\setcounter{figure}{0}
\section{Energy of isolated charge}

The electrostatic potential for an isolated charge is also governed by the Poisson equation, Eq.(3), with the boundary conditions for the potential to decay to zero at infinity,$\displaystyle{\lim_{\vec{r} \to \infty} V(\vec{r}) = 0}$. This makes the explicit solution of the Poisson equation by discretization of the Laplacian operator not tractable. A substantial number of modern approaches to the electrostatic problem under open boundary conditions, like the fast multipole method (FMM),~\cite{RN563} are instead based on direct summation of the potential induced by discretized charge elements, with some techniques utilized for improving efficiency.~\cite{RN564} In our case, the inhomogeneous dielectric profile complicates the problem, so we resort to a direct summation technique for the potential computation, as described next.

The approach we implement here is based on the image charge method. The key idea is that, for a discrete representation of the defect charge on the boundary of two dielectric media, the potential induced by the point charge elements on both sides of the dielectric boundary can be calculated analytically.~\cite{RN565} The situation is illustrated in Figure B1, which shows a schematic view of the slice of the charge distribution along the \textit{xz} plane. The thick black line denotes the boundary between two media with dielectric constants $\epsilon_1$ and $\epsilon_2$, respectively. For the potential computation an auxiliary grid is introduced, since the 1/$r$ Coulomb potential is singular; this auxiliary grid is shifted by a vector ($\frac{\Delta x}{2}$,$\frac{\Delta y}{2}$,$\frac{\Delta z}{2}$) compared to the charge mesh. For each charge point the contributions to all points on the potential grid are computed; there are two types of potential expression depending on the positions of charge and potential mesh points relative to the interface. For points on the same side of the interface, the potential is induced by the charge itself: 1/$\epsilon_1 r_1$, with the dielectric constant $\epsilon_1$ corresponding to the material in that part of the simulation domain. Another contribution is from the \enquote{image}charge, which induces a potential with effective screening factor $\frac{\epsilon_1 - \epsilon_2}{\epsilon_1 + \epsilon_2}$. In the limit of charge in vacuum near the metal surface, the effective screening factor is --1, which corresponds to the well-known limit of an image charge of equal magnitude and opposite sign. The lateral positions of the image charge are the same as those of the original charge element, and the $z$ coordinate is obtained by applying a mirror reflection operation in the plane separating the two media. For points on the opposite side of the interface, only the original charge element has a contribution with effective dielectric constant $\frac{2}{\epsilon_1 + \epsilon_2}$.

\begin{figure}
\centering
\includegraphics[scale=0.6]{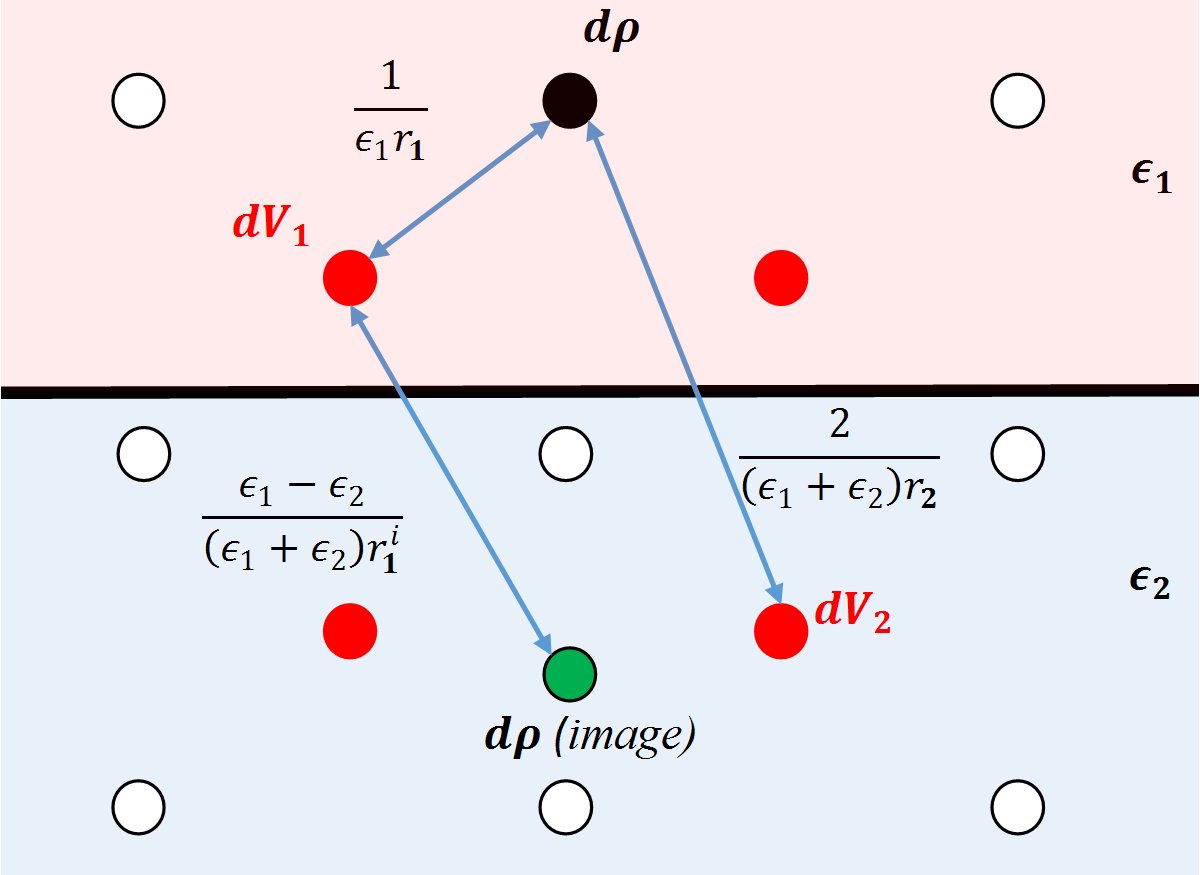}
\caption{Illustration of the computation under open boundary conditions. Points on the mesh for discretizing charge are shown as hollow white circles, and points of the potential mesh are shown as red circles.}
\end{figure}\label{fig:B1}

After obtaining the potential values on the offset grid the values of the potential are interpolated back on the original charge mesh. Due to the two iterations over all mesh points, the resulting computational cost is O($N_x^2$$N_y^2$$N_z^2$ ), with a much smaller contribution for the interpolation. However, it lends itself naturally to parallelization, where the computation of sub-arrays of the offset potential grid can be distributed among processes.

This approach has three major drawbacks: first, the method scales as the 6$^{th}$ power of mesh linear size; second, a single plane is a very crude approximation to the actual dielectric interface on the atomic scale; third, such a boundary model accommodates only one interface, thereby excluding 2D materials from consideration. We investigate another approach to computing the electrostatic energy under open boundary conditions through extrapolation of the periodic boundary conditions energy to infinite cell size. This method is inspired by the \enquote{scaling relationships} discussed in earlier methodology work,~\cite{RN24} where it was shown that the error in electrostatic energy scales as inverse of the supercell size. This method was mentioned in the literature before,~\cite{RN27} but it has two important caveats which we discuss here for the first time. 

The isolated energy can be recovered by carrying out a series of model electrostatic calculations for increasingly larger model supercells scaled by an integer factor $\alpha$ compared to the original size, and then fitting the resulting energies to a straight line as a function of 1/$\alpha$; the limit of 1/$\alpha \rightarrow 0$ is the electrostatic energy of an isolated charge. We consider a Gaussian charge of width 1.0 Bohr in vacuum in a cell of 12~$\times$~12~$\times$~12 Bohr. The extrapolation procedure is carried out by computing the electrostatic energies for that charge for a number of scaling factors up to 7 (84~$\times$~84~$\times$~84 Bohr), and fitting the resulting energies to a straight line. The result matches closely the true electrostatic self-energy of an isolated Gaussian charge distribution in vacuum $E_{Gauss}$ = 7.67 eV; the errors in extrapolated energy are 0.05, 0.03, 0.02 eV for maximal scaling factors of 3, 5, 7, respectively. The electrostatic energies and, correspondingly, differences between them scale inversely with the dielectric constant of the system, so the calculations in vacuum represent an upper bound on the error estimates in our case. Therefore, in practice it should be sufficient to set the scaling factor to 4 or 5. The electrostatic correction decreases with the increase in the size of the original supercell, so for larger systems the error is dominated by DFT errors in $E$($\alpha$) values, which depend on the sampling of the defect charge state, especially for anisotropic wavefunctions, and can reach $\sim$0.03 eV.

Another important component of the problem is the initial shape of the cell containing the charge. We have found that even for a Gaussian charge in vacuum any deviation of the original cell shape from cubic will result in very large errors (up to 5 eV for starting shape of 24~$\times$~24~$\times$~12 Bohr). This is a critical point that is rarely if ever mentioned in discussions of the extrapolation, and only for a cubic shape of the original supercell does the extrapolated energy converge to the proper limit.

\begin{acknowledgements}
DV, EK acknowledge support from ARO MURI W911NF-14-1-0247. DV, CMF acknowledges support from NSF CHE 1362616 Award in the Catalysis program. MGS acknowledges support from the Scientific and Technological Research Council of Turkey (TUBITAK) 2214-A Program, Grant no. 1059B141500480. Computational resources were provided by XSEDE (Grant No. TG-DMR120073), which is supported by NSF Grant No. ACI-1053575, and the Odyssey cluster, supported by the FAS Research Computing Group at Harvard University.
\end{acknowledgements}

\clearpage
\bibliography{references} 

\end{document}